\documentclass{article}
\usepackage{spconfa4,amsmath,graphicx}

\usepackage{amsfonts}
\usepackage{algorithmic}
\usepackage{algorithm}
\usepackage{array}
\usepackage[caption=false,font=normalsize,labelfont=sf,textfont=sf]{subfig}
\usepackage{textcomp}
\usepackage{stfloats}
\usepackage{url}
\usepackage{verbatim}
\usepackage{graphicx}
\usepackage[sort]{cite}
\usepackage{color}
\usepackage{xcolor}
\usepackage{tabularray}
\usepackage{tikz}
\usetikzlibrary{positioning}
\usepackage{adjustbox}
\usepackage{enumitem}   
\usepackage{pbalance}

\usepackage{booktabs}
\usepackage{hhline}
\usepackage{multirow}
\usepackage{makecell}
\usepackage{diagbox}
\usepackage{svg}
\usepackage{siunitx}
\usepackage[hidelinks]{hyperref} 

\usepackage[nolist]{acronym} 

\begin{acronym}
    \acro{df}[DF]{directivity factor}
    \acro{drr}[DRR]{direct-to-reverberant ratio}
    \acro{ndf}[NDF]{neural directional filtering}
    \acro{pesq}[PESQ]{perceptual evaluation of speech quality}
    \acro{rir}[RIR]{room impulse response}
    \acro{rtf}[RTF]{room transfer function}
    \acro{sdr}[SDR]{signal-to-distortion ratio}
    \acro{CDR}[CDR]{coherent-to-diffuse ratio}
    \acro{stft}[STFT]{short-time Fourier transform}
    \acro{vdm}[VDM]{virtual directional microphone}
    \acro{wpe}[WPE]{weighted prediction error}
    \acro{FBF}[FBF]{fixed beamformer}
	\acro{DNN}[DNN]{deep neural network}

	\acro{sdri}[$\Delta$SDR]{improvement in \ac{SDR} over the unprocessed signal}
	\acro{DMA}[DMA]{differential microphone array}
	\acro{DNN}[DNN]{deep neural network}
	\acro{DOA}[DOA]{direction-of-arrival}
	\acrodefplural{DOA}[DOAs]{directions-of-arrival}
	
	\acro{iSTFT}[iSTFT]{inverse short-time Fourier transform}
	\acro{CDMA}[CDMA]{circular \ac{DMA}}

	\acro{LDMA}[LDMA]{linear \ac{DMA}}
        \acro{LS}{least-squares}
	\acro{LSTM}[LSTM]{long short-term memory}
        \acro{BiLSTM}[BiLSTM]{bidirectional LSTM}
        \acro{UniLSTM}[UniLSTM]{unidirectional LSTM}
        \acro{WNG}[WNG]{white noise gain}
	\acro{RIRs}[RIRs]{room impulse responses}
	\acro{RIR}[RIR]{room impulse response}
        \acro{RTF}[RTF]{room transfer function}
        \acro{RTFs}[RTFs]{room transfer functions}        
	\acro{DPIR}[DPIR]{direct-path impulse response}
        \acro{MVDR}[MVDR]{minimum variance distortionless response}
        \acro{LCMV}[LCMV]{linear-constraint minimum-variance}
        \acro{PMWF}[PMWF]{parametric multichannel wiener filter}
        \acro{GSC}[GSC]{Generalized sidelobe canceller}
        \acro{FT-JNF}[FT-JNF]{joint spatial and temporal-spectral non-linear filtering}  
        \acro{JNF}[JNF]{joint non-linear filtering }        
        \acro{SSF}[SSF]{spatially selective
deep non-linear filter } 
	\acro{SDR}[SDR]{signal-to-distortion ratio}
        \acro{noisySDR}[reference microphone]{\ac{SDR} of the unprocessed omnidirectional reference microphone}
    \acrodefplural{noisySDR}[$\textrm{SDRs}^\textrm{omni}$]{\acp{SDR} of the unprocessed omnidirectional microphone}
	\acro{SNR}[SNR]{signal-to-noise ratio}
	\acro{STFT}[STFT]{short-time Fourier transform}
         \acro{MAE}[MAE]{mean absolute error}	
	\acro{TF}[TF]{time-frequency}
	\acro{tsdr}[SA-$\varepsilon$-tSDR]{source-aggregated and regularized thresholded \ac{SDR}}
      \acro{STOI}[STOI]{short term objective intelligibility}
    \acro{PESQ}[PESQ]{perceptual evaluation of speech quality}
	
	\acro{UCA}[UCA]{uniform circular array}
        \acro{NDF}[NDF]{neural directional filtering} 
        \acro{SHONDC}[SHONDC]{steerable high-order neural directional coding}
        \acro{NDSC}[NDSC]{neural directional speech coding}
        \acro{NDC}[NDC]{neural directional coding}
        \acro{WNG}[WNG]{white noise gain}
        \acro{DF}[DF]{directivity factor}
        \acro{DI}[DI]{directivity index}
        \acro{HRTF}[HRTF]{head-related transfer function}
        \acro{ILD}[ILD]{interaural level difference}
        \acro{FiLM}[FiLM]{feature-wise linear modulation}     
        \acro{PESQ}[PESQ]{perceptual evaluation of speech quality}
        \acro{UNDF}[UNDF]{neural directional filtering with user-defined directivity patterns} 
    \acro{VDM}[VDM]{virtual directional microphone}
    \acro{DirAC}[DirAC]{directional audio coding}
    \acro{FOA}[FOA]{first-order ambisonics}
    \acro{HOA}[HOA]{high-order ambisonics}
        \acro{ATF}[ATF]{acousitc transfer function}

        \acro{WNG}[WNG]{white noise gain}
        \acro{DF}[DF]{directivity factor}
        \acro{DI}[DI]{directivity index}
        \acro{HRTF}[HRTF]{head-related transfer function}
        \acro{ILD}[ILD]{interaural level difference}
        \acro{FiLM}[FiLM]{feature-wise linear modulation}
        \acro{NDBF}[NDBF]{high-order frequency-invariant neural beamforming}
        \acro{DPTF}[DPTF]{direct-path transfer function}
        \acro{CMA}[CMA]{circular microphone array}
        \acro{LMA}[LMA]{linear microphone array}
        \acro{NDBF}[NDBF]{steerable high-order neural beamformer}
        \acro{SI-SDR}[SI-SDR]{scale-invariant signal-to-distortion ratio}
        \acro{DBF}[DBF]{differential beamformer}
       \acro{NDBF}[NDBF]{neural differential beamformer}

\end{acronym}

\title{Dual-Microphone Steerable High-Order Neural Differential Beamformer}
\name{Weilong Huang, Emanu{\"e}l A. P. Habets}
\address{International Audio Laboratories Erlangen\textsuperscript{$\ast$}, Am Wolfsmantel 33, 91058 Erlangen, Germany\thanks{\textsuperscript{$\ast$}A joint institution of Fraunhofer IIS and Friedrich-Alexander-Universit{\"a}t Erlangen-N{\"u}rnberg (FAU), Germany. The authors gratefully acknowledge the scientific support and HPC resources provided by the Erlangen National High Performance Computing Center (NHR@FAU) of the Friedrich-Alexander-Universität Erlangen-Nürnberg (FAU). The hardware is funded by the German Research Foundation (DFG).}}
\begin{document}
\ninept
\maketitle

\begin{abstract}
Linear arrays of omnidirectional microphones produce beampatterns that are symmetric about the array axis. For these arrays, a beamformer is considered steerable if its beampattern maintains the same shape in the semicircular plane across all look directions from 0° to 180°. For dual-microphone arrays, conventional differential beamformers are generally non-steerable and restricted to first-order, which significantly limits spatial selectivity. To address these limitations, this study presents a neural differential beamformer (NDBF) with a dual-microphone array. The contributions are as follows: (i) NDBF is steerable; (ii) NDBF achieves high-order frequency-invariant beampatterns; and (iii) NDBF enables stereo recording using only two closely spaced omnidirectional microphones. Experimental results demonstrate that NDBF outperforms existing methods while overcoming the limitations of classical differential beamforming. 
\end{abstract}

\section{Introduction}
Recently, differential beamformers for linear arrays have received significant attention in \ac{DMA} research \cite{benesty2012study, chen2014design, 9261932}. The largest order of such beamformers is $Q-1$, where $Q$ denotes the number of microphones \cite{benesty2012study}. A dual-microphone array, which constitutes the minimum number of microphones for a linear array, restricts the \ac{DMA} to first order and substantially limits spatial filtering capability.

The steerability of beamformers for a circular array is readily attainable and has been extensively studied \cite{10138024,9980184,steerable_NDF_FA_2025}. In contrast, most differential beamformers designed for a linear array are generally assumed to have a look direction at $0^{\circ}$ (endfire) \cite{benesty2012study,chen2014design}, as this configuration typically yields the highest \ac{DF}. However, the ability to steer the look direction away from $0^{\circ}$ while maintaining the same beampattern is essential for practical applications of a linear array, including televisions, tablets, and laptops. For a linear array, ``steerability" refers to the ability of a beamformer to maintain a consistent beampattern across look directions from $0^{\circ}$ to $180^{\circ}$ in the semicircular plane, as described in \cite{9261932}. 

Two situations fall short of full steerability: (1) \textit{partially steerable}, where the beampattern achieves 0~\unit{\decibel} at the look direction and does not exceed 0~\unit{\decibel} elsewhere, but varies across look directions; and (2) \textit{non-steerable}, where the beampattern achieves 0~\unit{\decibel} at the look direction, but exceeds 0~\unit{\decibel} in some other directions \cite{9261932}. For a differential beamformer with a linear array, achieving steerable beampatterns is feasible only with high-order designs and requires that the null positions satisfy specific conditions \cite{9261932}. Therefore, a first-order differential beamformer with dual microphones is hardly steerable. Recent linear superarrays \cite{10103666, 10542464, huang2025robust,11447415} enable differential beamformers to achieve steerability, but their core concept relies on a hybrid usage of omnidirectional and directional microphones. Therefore, developing a steerable differential beamformer for a linear array, particularly with only two omnidirectional microphones, remains an open research problem.

Most \ac{DNN}-based methods have been proposed for performing spatial filtering within an angular region \cite{deep_zoom, wen2025neural} or directly focusing on speech separation or extraction \cite{ftjnf, mcnet}, where the beampattern is not explicitly controlled. Recently, \ac{NDF} \cite{ndf_iwaenc} proposed to estimate a single-channel mask for nonlinear spatial filtering based on a predefined beampattern using a circular array. This approach paves the way for beampattern-oriented neural spatial filtering and has the potential to address the limitations in differential beamforming. However, \ac{NDF} has so far been studied only for circular arrays, and its single-channel masking strategy may not fully exploit the spatial information available for beamforming, particularly for dual-microphone arrays, where spatial degrees of freedom are very limited. As an extension of \ac{NDF} to beamforming for a dual-microphone array, we propose a \ac{NDBF} that, using a \ac{DNN}, produces beampatterns similar to those of classical \acp{DMA}. Experimental results demonstrate that \ac{NDBF} with a dual-microphone array achieves steerable high-order beampatterns and surpasses existing methods. Leveraging the steerability and beampattern-oriented spatial filtering of NDBF, we demonstrate that stereo recording can be achieved with only two closely spaced omnidirectional microphones.

\begin{figure}[t!] 
\centering	
\includegraphics[width=0.899\linewidth]{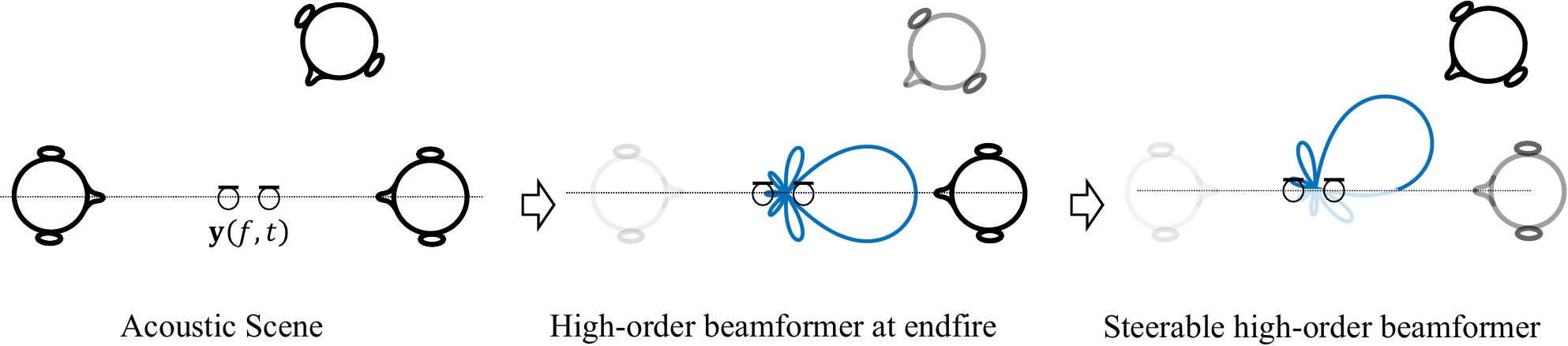}
	\caption{The goal of the steerable high-order neural differential beamforming with dual-omnidirectional microphones.}
	\label{fig: diagram} \vspace{-1.5em}
\end{figure}

\section{Problem Formulation}
We consider a dual-microphone array comprising two omnidirectional microphones to capture $N$ sound sources in an anechoic environment. The corresponding microphone array signals in \ac{STFT} domain are denoted by $\textbf{y}(f,t) = [Y_1(f,t), Y_2(f,t) ]$, where $f$ and $t$ denote the frequency and time indices, respectively. The mixture signal on the first microphone $Y_1(f,t)$ can be decomposed as
\vspace*{-0.15cm}
\begin{equation} \label{eqn:mic_sig}
        Y_1(f,t) = \sum_{n=1}^{N} X_{1,n}(f,t) + V_1(f,t),
\end{equation}
where $V_1(f,t)$ is spatially uncorrelated sensor noise and $X_{1,n}(f,t) = H_{\mathbf{p}_1,n}(f) \, X_{n}(f,t)$, where $H_{\mathbf{p}_1,n}(f)$ models the \ac{DPTF} between the $n$-th source $X_{n}(f,t)$ and the first microphone at position $\mathbf{p}_1$.

For this dual-microphone array, the goal of the \ac{NDBF} is to capture the $N$ sound sources using a steerable high-order \ac{DMA} beampattern (see Fig.~\ref{fig: diagram}). This beampattern is denoted by $\Lambda_{\theta_\textrm{s}}(\theta_\textrm{n})$, where ${\theta_\textrm{s}}$ is the steering direction of the beampattern and $\theta_\textrm{n}$ denotes the \ac{DOA} of the $n$-th source with respect to the position $\mathbf{p}_\textrm{c}$ (the center of the dual-microphone array). The target signal of the \ac{NDBF} is expressed as
    \vspace*{-0.15cm}
    \begin{equation}\label{eqn:vdm_sig} 
        Z_{\theta_\textrm{s}}(f,t) = \sum_{n=1}^{N} \Lambda_{\theta_\textrm{s}}(\theta_\textrm{n}) \, H_{{\mathbf{p}_{\textrm{c}}},n}(f) \, X_n(f,t),
    \end{equation} 
where $H_{{\mathbf{p}_{\textrm{c}}},n}(f)$ models the \ac{DPTF} between the $n$-th source $X_{n}(f,t)$ and the position $\mathbf{p}_\textrm{c}$. In this work, we propose a \ac{DNN} based method to obtain the beamformer weights $\textbf{w}_{\theta_\textrm{s}}(f)$ given any steering direction ${\theta_\textrm{s}}$ at inference. 

\section{Proposed Method}
\begin{figure}[t!] 
\centering	
\includegraphics[width=0.759\linewidth]{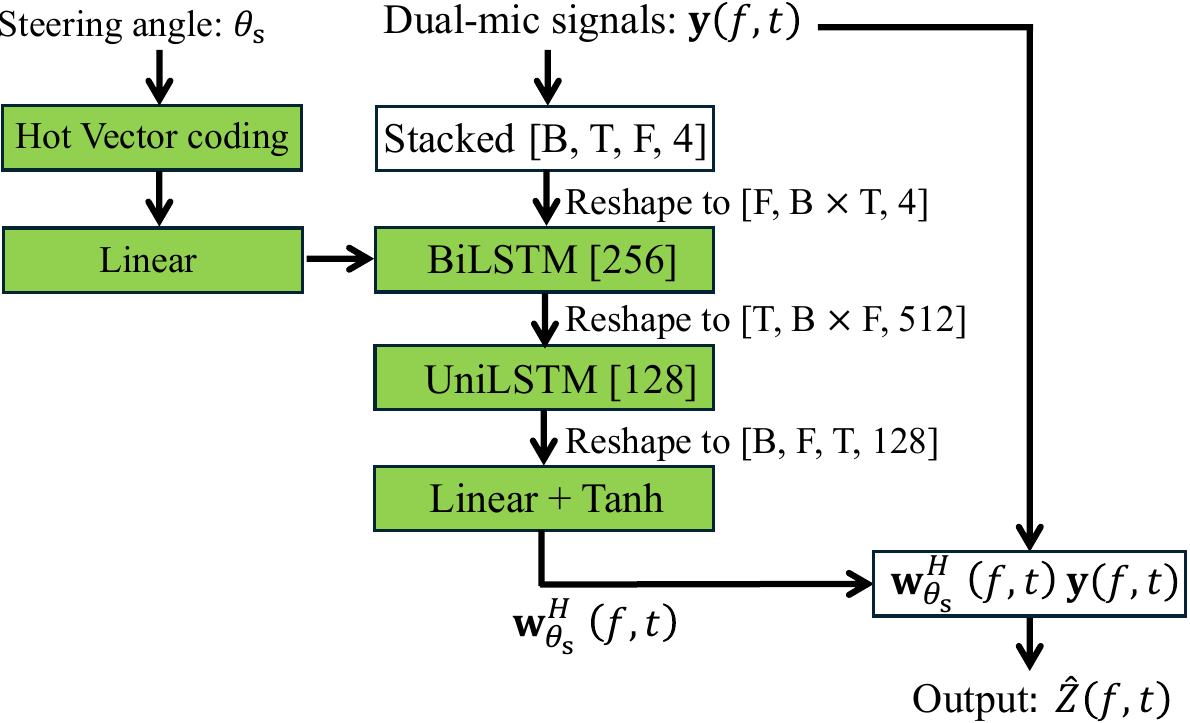}
	\caption{Extended version of the JNF-SSF architecture \cite{tesch2023multi}, which outputs two complex weights per time $t$ and frequency $f$.}
	\label{fig: dnn}
    \vspace*{-0.15cm}
\end{figure}
\subsection{DNN Architecture and Loss Function}
The JNF-SSF \cite{tesch2023multi} is adopted as \ac{DNN} architecture for the \ac{NDBF} task, with a modification in which the single-channel complex mask is replaced by a vector of complex weights at the output of the final linear layer (see the left side in Fig.~\ref{fig: dnn}). The real and imaginary components of the dual-microphone signals in the \ac{STFT} domain are stacked along the channel dimension, resulting in an input with dimensions of $[B, T, F, 4]$, where $B$ indicates the batch size, $T$ represents the number of time frames, and $F$ denotes the number of frequency bins. The stacked inputs are processed by two distinct \ac{LSTM} modules. The first \ac{LSTM} is bidirectional and operates along the frequency dimension, thereby modeling instantaneous spectro-spatial relationships in the input \cite{tesch_insights}. Its output is subsequently processed by a unidirectional \ac{LSTM} module that captures temporal relationships, treating the frequency dimension as the batch dimension and thus modeling all frequencies independently \cite{tesch_insights}. As in \cite{tesch2023multi}, the desired look direction of the beampattern is encoded as a one-hot vector, which is mapped to an embedding of the same dimension as the first \ac{LSTM} hidden states via a linear layer. The output of this layer is then used to initialize the \ac{LSTM}'s states for each time frame. Finally, a linear layer with a hyperbolic tangent activation function computes the \ac{NDBF} weights $\textbf{w}_{\theta_\textrm{s}}(f)$. These weights are applied to the microphone array signals $\textbf{y}(f,t)$ to obtain an estimate of the target signal, i.e., $\widehat{Z}_{\theta_\textrm{s}}(f,t) =\textbf{w}_{\theta_\textrm{s}}^{H}(f) \ \textbf{y}(f,t)$.

The training loss is based on a batch-aggregated normalized $\mathcal{L}_{\textrm{1}}$-loss function \cite{steerable_NDF_FA_2025, huang2025neural-journal,huang2025neural}, expressed as: 
\vspace*{-0.15cm}
\begin{equation}\label{eqn:loss_func}
\mathcal{L}_{\textrm{1}}=\frac{\sum_{b=1}^B \left \lVert \mathbf{z}^{b} - \hat{\mathbf{z}}^{b} \right \rVert_{1}}{ \sum_{b=1}^B \left \lVert  \mathbf{z}^{b} \right \rVert_{1} + \epsilon},
\end{equation}
where the time-domain signals $\hat{z}$ and ${z}$ correspond to \ac{STFT} representations $\widehat{Z}$ and ${Z}$, respectively. 

\subsection{Training Strategy}

\begin{figure}[t!] 
\centering	
\includegraphics[width=0.899\linewidth]{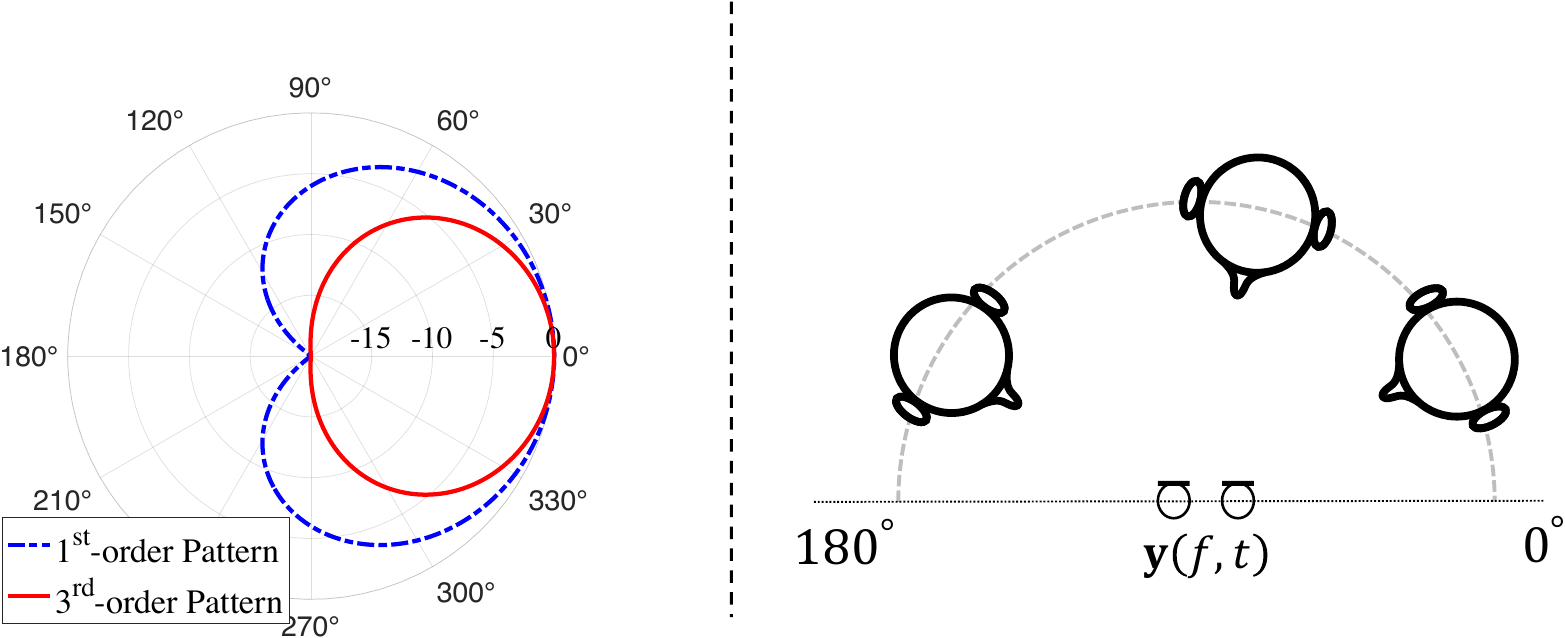}
	\caption{Target beampatterns (left side) and a source-array setup example (right side).  }
	\label{fig: patterns}
        \vspace*{-0.15cm}
\end{figure}

A $J$th-order \ac{DMA} beampattern \cite{elko2000superdirectional} is defined as follows $\Lambda_{\theta_\textrm{s}} =\sum_{j=0}^{J} a_{j} \, \cos^{j}(\theta - \theta_\textrm{s})$, where $\theta_\textrm{s}$ represents the steering direction of the pattern, while $a_{j}$, where $j \in \{0, 1, \ldots, J \}$, are real-valued coefficients that define the pattern's shape, particularly affecting the sidelobes. In sound capture scenarios, primary attention is given to controlling the mainlobe of the patterns. Therefore, a simplified \ac{DMA} pattern is considered as follows:
\vspace*{-0.15cm}
\begin{equation}\label{eqn:dma_new}
\Lambda(\theta) = (\mu +  (1-\mu) \cos(\theta - \theta_\textrm{s}) )^{J},
\end{equation}
where $\mu$ is a real-valued parameter within $[0, 1]$ that determines the null position. The patterns described by \eqref{eqn:dma_new} emphasize mainlobe shape, which is governed by the order $J$ and $\mu$. In this paper, two example beampatterns were employed as the target beampatterns in Fig.~\ref{fig: patterns}. The first target pattern is a $1^{\textrm{st}}$-order Cardioid with $\mu = 0.5, J = 1$. The second is a $3^{\textrm{rd}}$-order Cardioid pattern with $\mu = 0.5, J = 3$. 

To learn the \ac{DMA} beampattern with a linear array in the free field, the sound sources are positioned along a semicircle, with the microphone array located along the $0^{\circ}$-$180^{\circ}$ axis and being concentric with the semicircle (an example is shown in Fig.~\ref{fig: patterns}). To simulate a source-array setup with $N$~source positions, $N$~source positions are randomly selected on the semicircle, and \acp{DPTF} are simulated for all microphones and sources using the RIR generator \cite{RIRGenerator} with a reflection order of zero. Microphone array signals are then obtained using \eqref{eqn:mic_sig}. For each source-array setup, we simulate $M$ target signals for look directions uniformly spanning $0$ to $180$ degrees, resulting in a steering angular resolution of $\frac{180^{\circ}}{M}$. The $m$-th target signal $Z_{\theta^{m}_{\textrm{s}}}[f,t]$ corresponding to the steering direction $\theta^{m}_{\textrm{s}}$ is obtained using \eqref{eqn:vdm_sig}. During training, we treat microphone-array signals from each acoustic scene, paired with a target signal, as one training sample.

\section{Experimental Setup}
\subsection{Datasets}
Speech signals from the `train-clean-360' and `dev-clean' subsets of the LibriSpeech corpus \cite{librispeech} served as source signals for training and validation, respectively. For the test sets, speech signals were selected from the EARS dataset \cite{richter2024ears} using a minimum loudness of $-42$~dBFS \cite{loudness}. The parameter $M$ was set to 36, resulting in a steering resolution of $5^{\circ}$. A total of 1440 random source-array setups were simulated, yielding $1440 \times 36$ samples for the training set. The test set contains 3240 samples. Each sample in all datasets has a duration of 4~seconds. In each sample, the concurrent sources were randomly drawn from the setup's source positions, with up to three for training and exactly two for testing. Source positions for training and validation were selected from $\theta_{\textrm{n}} \in \{0^{\circ}, 5^{\circ}, \ldots, 175^{\circ} \}$ and $\theta_{\textrm{n}} \in \{2.5^{\circ}, 7.5^{\circ}, \ldots, 177.5^{\circ} \}$, respectively. The source positions in test sets were selected from $\theta_{\textrm{n}} \in \{1.25^{\circ}, 3.75^{\circ}, \ldots, 178.75^{\circ} \}$. The distance between the two microphones was set to \qty{3}{\cm}, and the source-array distance for all speakers was fixed at \qty{1.5}{\m}. Microphone sensor noise was added with an SNR of $30$~dB. STFT settings and training details followed \ac{NDF} \cite{ndf_iwaenc}.

\subsection{Performance Measures}
\noindent \textbf{Estimated Beampattern}: For each test sample, the array signals for the $n$-th source is denoted by $\textbf{x}_{n}(f,t) = [X_{1,n}(f,t), X_{2, n}(f,t) ]$. Then, the estimated beamformer $\textbf{h}(f)$ is applied separately to the array signals for each source. Subsequently, the wideband power ratio $\xi[ \theta_\textrm{n}]$ for the $n$-th source is calculated as follows:
\vspace*{-0.2cm}
\begin{equation}\label{eqn:patternEstimation}
\xi[ \theta_\textrm{n}] = \frac{ \sum_{f=1}^{F}\sum_{t=1}^{T} \left| \textbf{h}^{H}(f) \textbf{x}_{n}(f,t) \right|^2}{\sum_{f=1}^{F}\sum_{t=1}^{T}\left| X_{1, n}[f, t] \right|^2}.
\end{equation}
The arithmetic mean of the ratios ($\xi[ \theta_\textrm{n}]$) is then computed over all test samples from the same direction $\theta_\textrm{n}$ to obtain the final estimated wideband beampattern. Similarly, by removing the frequency-based summation from both the numerator and denominator in \eqref{eqn:patternEstimation}, the narrowband beampattern is computed.

\noindent \textbf{SDR}: The \ac{SDR} \cite{vincent2006performance}, averaged over the test set, is used to measure the distortion in the estimated signals compared to the target signals.

\section{Experimental Results}
\subsection{Performance Analysis}
\begin{table}[t!]
  \centering
  \vspace{-6pt}
  \caption{SDR (\si{\decibel}) performance: comparison of proposed method with baseline for the end-fire direction.}
  \label{tab: sdr_film}
  \resizebox{.4\textwidth}{!}{
    \begin{tabular}{l rr}
    \toprule
       \multicolumn{1}{c}{\textbf{Method}} & 
       \multicolumn{1}{c}{\textrm{$1^{\textrm{st}}$-order Pattern}} & 
       \multicolumn{1}{c}{$3^{\textrm{rd}}$-order Pattern}  \\
       \midrule
       DMA \cite{benesty2012study} & -0.99  &  N/A   \\
       Parametric spatial filter \cite{thiergart2014informed} & 13.77  &  10.32 \\
       NDF \cite{ndf_iwaenc} & 25.85  & 23.13\\
       Proposed NDBF  & \bfseries 25.93 & \bfseries 23.24  \\ 
      \bottomrule
    \end{tabular}%
   }
  \vspace{-1.5em}
\end{table}


\begin{figure}[t!]
    \vspace{-1em}
    \begin{minipage}[b]{0.458\linewidth} 
        \centering
        \includegraphics[width=\linewidth]{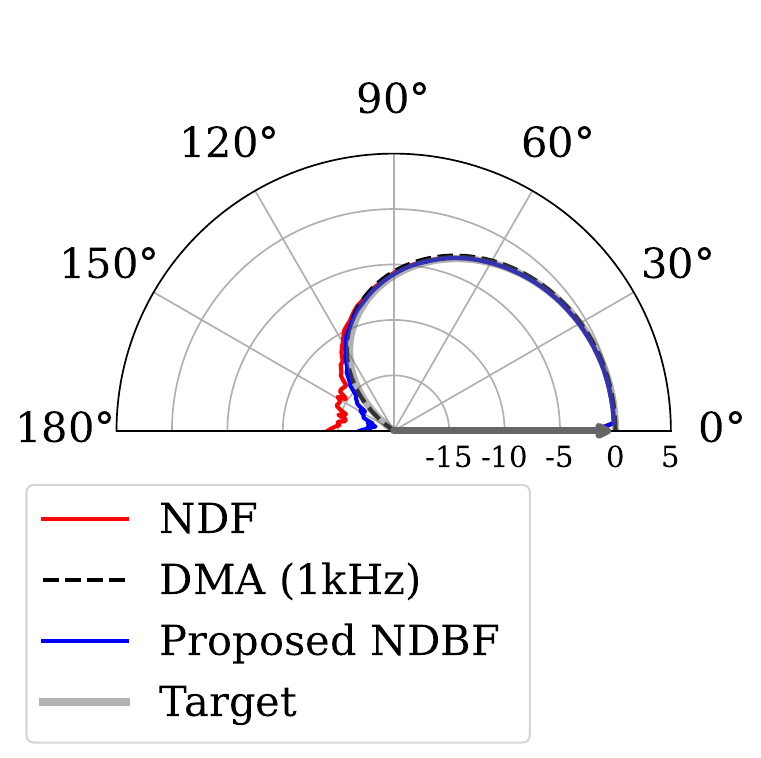}
        (a) $\theta_{\textrm{s}} = 0^\circ$, $1^{\textrm{st}}$-order.
    \end{minipage}
    \begin{minipage}[b]{0.458\linewidth} 
        \centering
        \includegraphics[width=\linewidth]{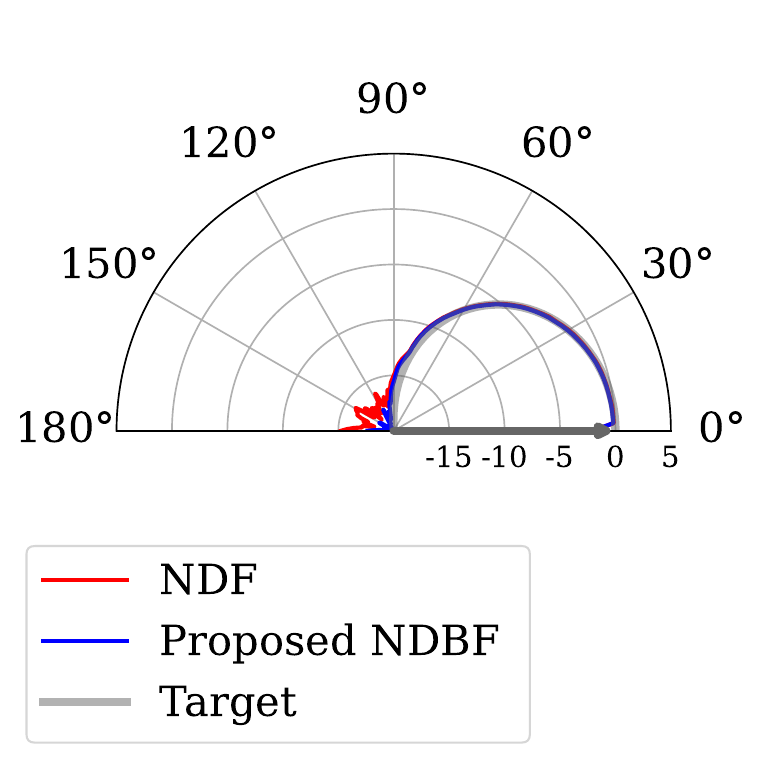}
        (b) $\theta_{\textrm{s}} = 0^\circ$, $3^{\textrm{rd}}$-order.
    \end{minipage}

    \begin{minipage}[b]{0.458\linewidth} 
        \centering
        \includegraphics[width=\linewidth, trim=0 30 0 0, clip]{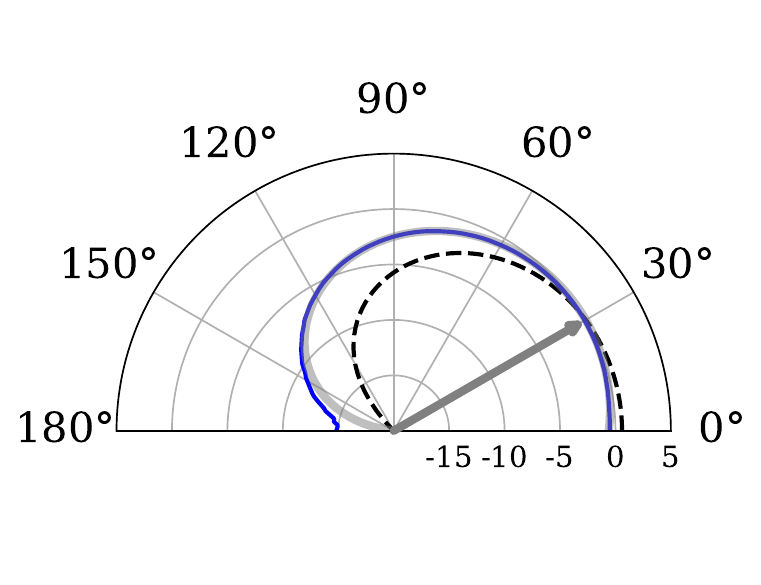}
        (c) $\theta_{\textrm{s}} = 30^\circ$, $1^{\textrm{st}}$-order.
    \end{minipage}
    \begin{minipage}[b]{0.458\linewidth} 
        \centering
        \includegraphics[width=\linewidth, trim=0 30 0 0, clip]{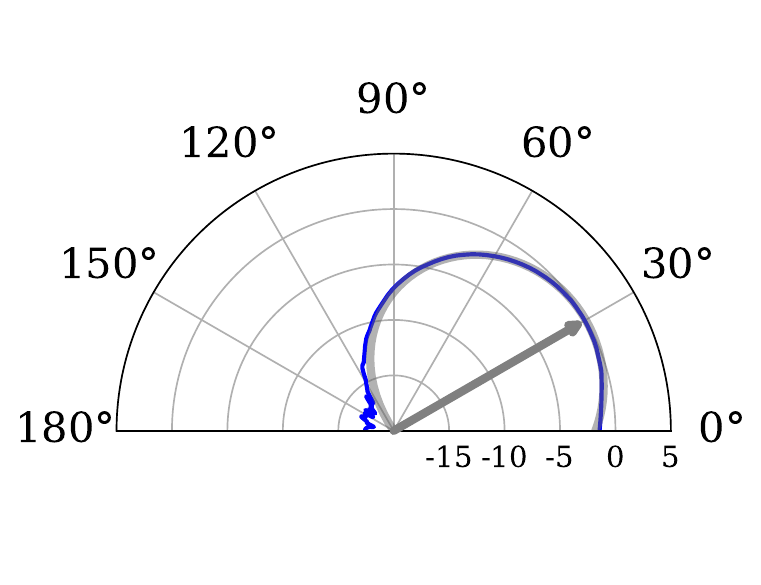}
        (d) $\theta_{\textrm{s}} = 30^\circ$, $3^{\textrm{rd}}$-order.
    \end{minipage}

    \begin{minipage}[b]{0.458\linewidth} 
        \centering
        \includegraphics[width=\linewidth, trim=0 30 0 0, clip]{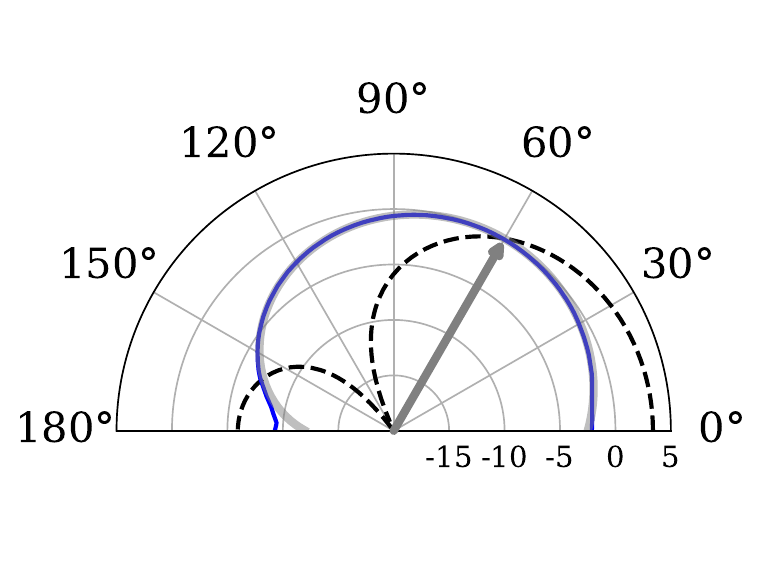}
        (e) $\theta_{\textrm{s}} = 60^\circ$, $1^{\textrm{st}}$-order.
    \end{minipage}
        \begin{minipage}[b]{0.458\linewidth} 
        \centering
        \includegraphics[width=\linewidth, trim=0 30 0 0, clip]{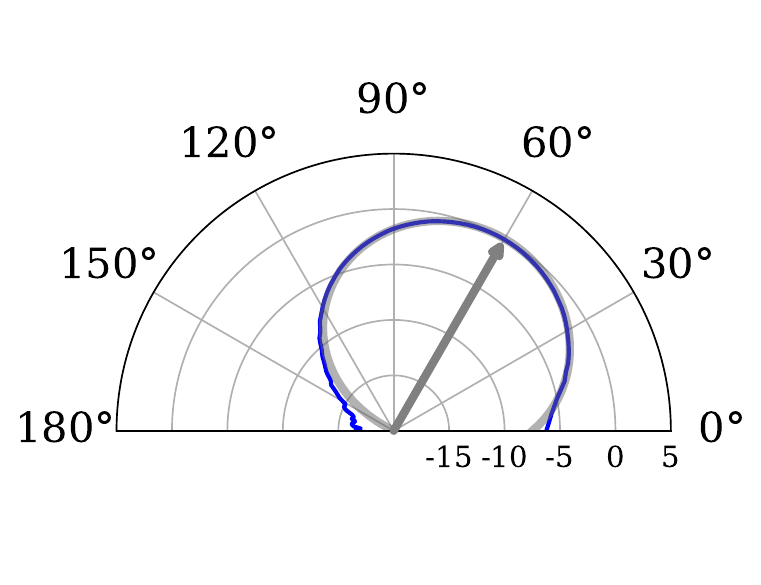}
        (f) $\theta_{\textrm{s}} = 60^\circ$, $3^{\textrm{rd}}$-order.
    \end{minipage}

    \begin{minipage}[b]{0.458\linewidth} 
        \centering
        \includegraphics[width=\linewidth, trim=0 30 0 0, clip]{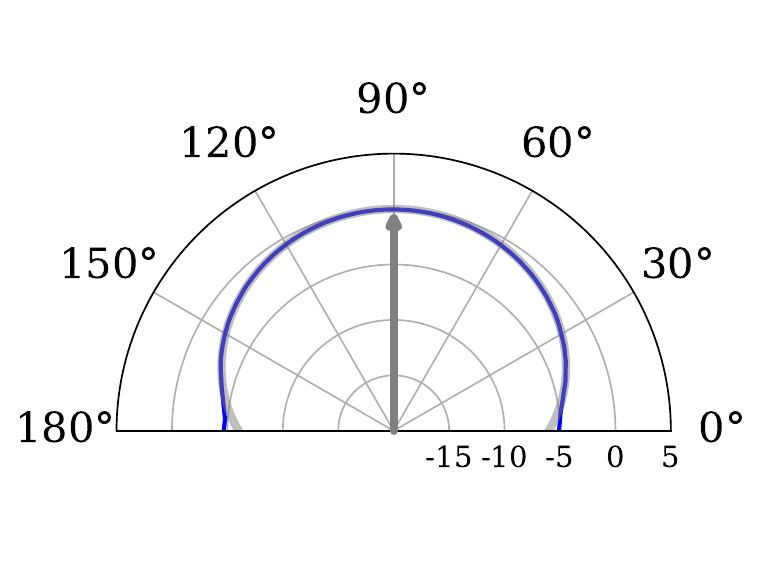}
        (g) $\theta_{\textrm{s}} = 90^\circ$, $1^{\textrm{st}}$-order.
    \end{minipage}
    \begin{minipage}[b]{0.458\linewidth} 
        \centering
        \includegraphics[width=\linewidth, trim=0 30 0 0, clip]{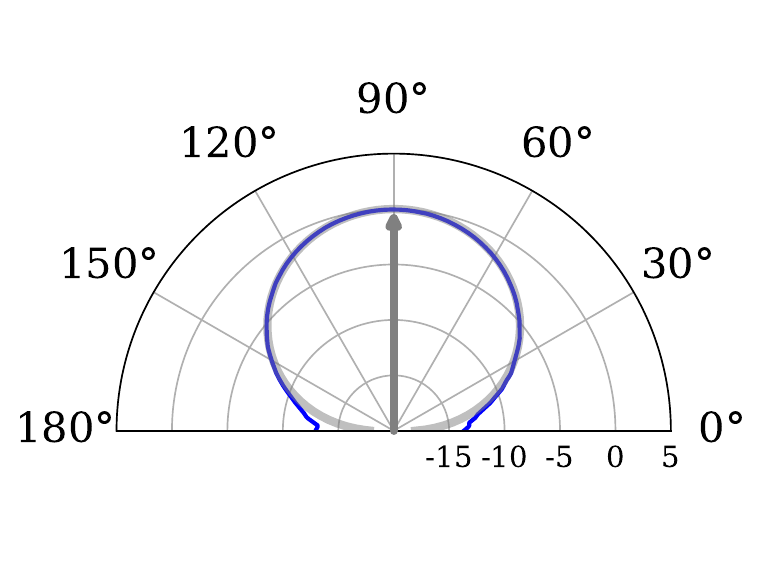}
        (h) $\theta_{\textrm{s}} = 90^\circ$, $3^{\textrm{rd}}$-order.
    \end{minipage}
       
	\caption{Steerability study of the NDBF for various look directions $\theta_{\textrm{s}}$ and compared with the baselines. } \vspace{-2em}
	\label{fig:1st-sbp}		
\end{figure}

\begin{figure}[t!]
    \begin{minipage}[b]{0.429\linewidth} 
        \centering
        \includegraphics[width=\linewidth]{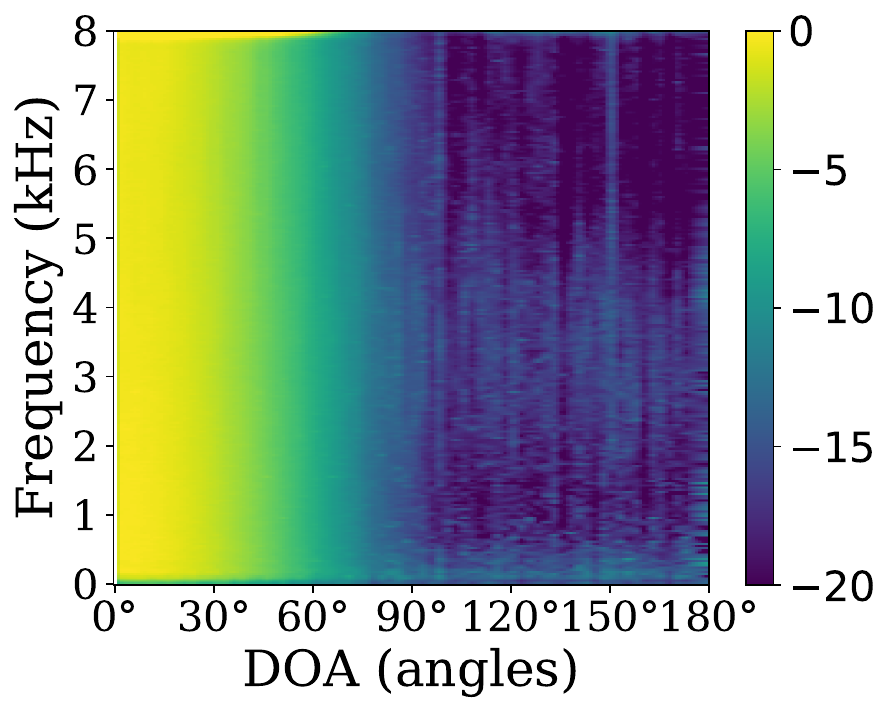}
        (a)  NDF \cite{ndf_iwaenc}, $\theta_{\textrm{s}}= 0^\circ$.
    \end{minipage}
    \begin{minipage}[b]{0.429\linewidth} 
        \centering
        \includegraphics[width=\linewidth]{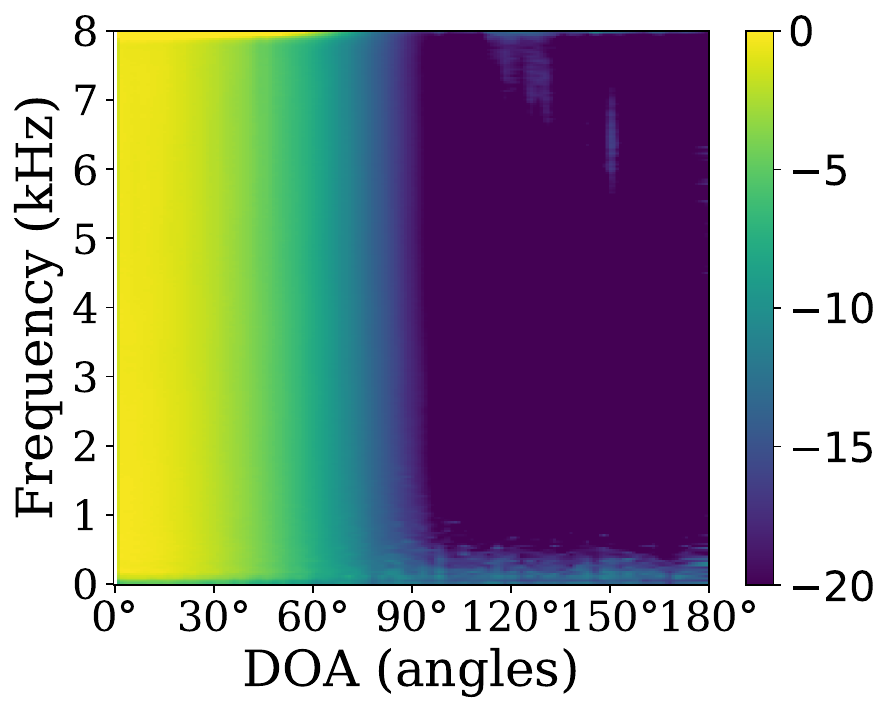}
        (b) NDBF, $\theta_{\textrm{s}}= 0^\circ$.
    \end{minipage}
    
    \begin{minipage}[b]{0.429\linewidth} 
        \centering
     \includegraphics[width=\linewidth]{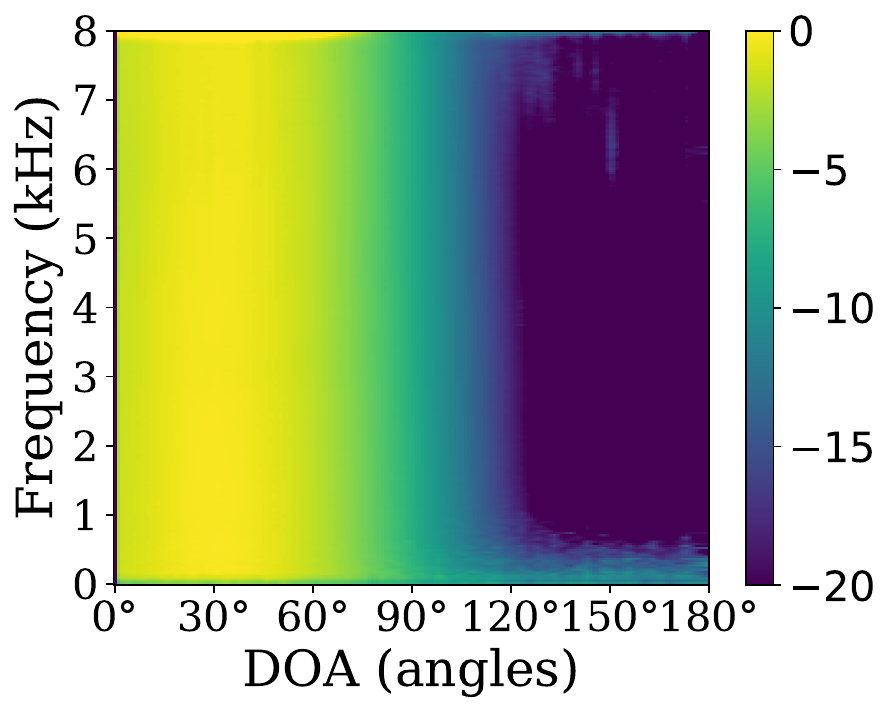}
        (c) NDBF, $\theta_{\textrm{s}}= 30^\circ$.
    \end{minipage}
    \begin{minipage}[b]{0.429\linewidth}
        \centering
        \includegraphics[width=\linewidth]{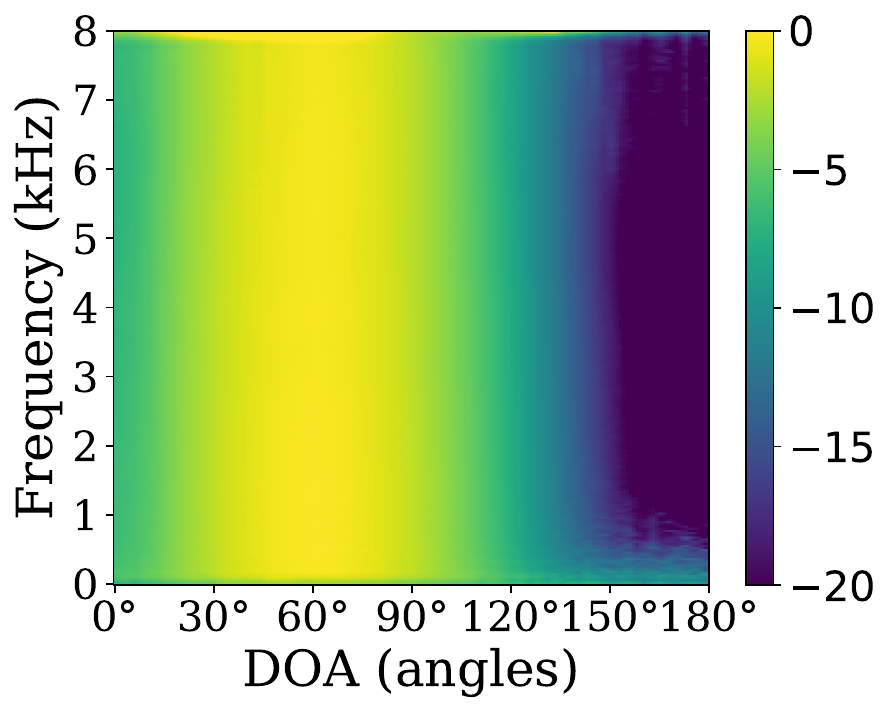}
        (d) NDBF, $\theta_{\textrm{s}}= 60^\circ$.
    \end{minipage}
   
	\caption{Estimated $3^{\textrm{rd}}$-order narrowband beampatterns. (a) Baseline NDF \cite{ndf_iwaenc} at $\theta_{\textrm{s}}= 0^\circ$. (b)-(d) Proposed NDBF at steering angles $\theta_{\textrm{s}}= 0^\circ$, $30^\circ$, and $60^\circ$.
     } 
	\label{fig:nbbp}		
\end{figure}

To the best of our knowledge, three methods can perform spatial filtering exactly based on a predefined beampattern: the classic \ac{DMA} \cite{benesty2012study}, the \ac{NDF} \cite{ndf_iwaenc}, and the parametric spatial filter \cite{thiergart2014informed}. For a fair comparison, \ac{NDF} was trained for a dual-microphone array using the same training strategy and dataset as the \ac{NDBF}. 

As shown in Table~1, the proposed NDBF achieves the highest SDR. The performance of the \ac{DMA} and the parametric spatial filter is significantly lower than that of the two neural network-based methods. The \ac{DMA} is subject to the well-known white-noise amplification problem at low frequencies\cite{benesty2016fundamentals} and spatial aliasing above 5.7~\unit{\kilo\hertz} due to an inter-microphone spacing of \qty{3}{\cm}, which substantially reduces its SDR. The parametric spatial filter is also impacted by spatial aliasing. To further compare NDF and the proposed NDBF, the estimated beampatterns produced by both methods were analyzed, as shown in Figs.~\ref{fig:1st-sbp}(a) and (b) (wideband beampattern) and Figs.~\ref{fig:nbbp}(a) and (b) (narrowband beampattern). Figs.~\ref{fig:1st-sbp}(a) and (b) demonstrate that both methods effectively approximate the mainlobe of the target beampattern, while \ac{NDBF} provides stronger suppression near the null position. This finding explains the SDR improvement of \ac{NDBF} over \ac{NDF} in Table~1, and also aligns with Figs.~\ref{fig:nbbp}(a) and (b), which show that \ac{NDBF} produces a cleaner suppression region in the narrowband beampattern. Additionally, both NDBF and NDF achieve a frequency-invariant beampattern without spatial aliasing for the broadband speech signals under test. The capability of NDF to achieve frequency-invariant beampattern has been previously analyzed in \cite{huang2025neural-journal, mannanova2025analysis}, and the present study confirms that the proposed NDBF also exhibits this property.



Since the \ac{DMA} does not guarantee a frequency-invariant beampattern, we analyze its beampattern at 1~\unit{\kilo\hertz} for comparison to investigate the steerability of the \ac{NDBF}. We present estimated wideband beampatterns of \ac{NDBF} for four different look directions (0°, 30°, 60°, and 90°) as illustrated in Fig.~\ref{fig:1st-sbp},  where (a), (c), (e), and (g) corresponds to the $1^{\textrm{st}}$-order target beampattern and (b), (d), (f), and (h) corresponds to the $3^{\textrm{rd}}$-order target beampattern. The \ac{NDBF} effectively approximates the target beampattern, and its beampattern remains consistent across different look directions, indicating that \ac{NDBF} is steerable. In contrast, the \ac{DMA} often achieves a response of 0~\unit{\decibel} in the target direction but exceeds 0~\unit{\decibel} in other directions, which is a typical non-steerable characteristic \cite{9261932}.
Figs.~\ref{fig:nbbp}(c) and (d) show the estimated $3^{\textrm{rd}}$-order narrowband beampatterns of the NDBF for look directions at $30^\circ$ and $60^\circ$. These results demonstrate that steerable high-order beampatterns maintain frequency-invariant properties similar to those observed in the endfire direction.

\subsection{Application to Stereo Recording}
\begin{figure}[t!] 
\centering	
\includegraphics[width=0.899\linewidth]{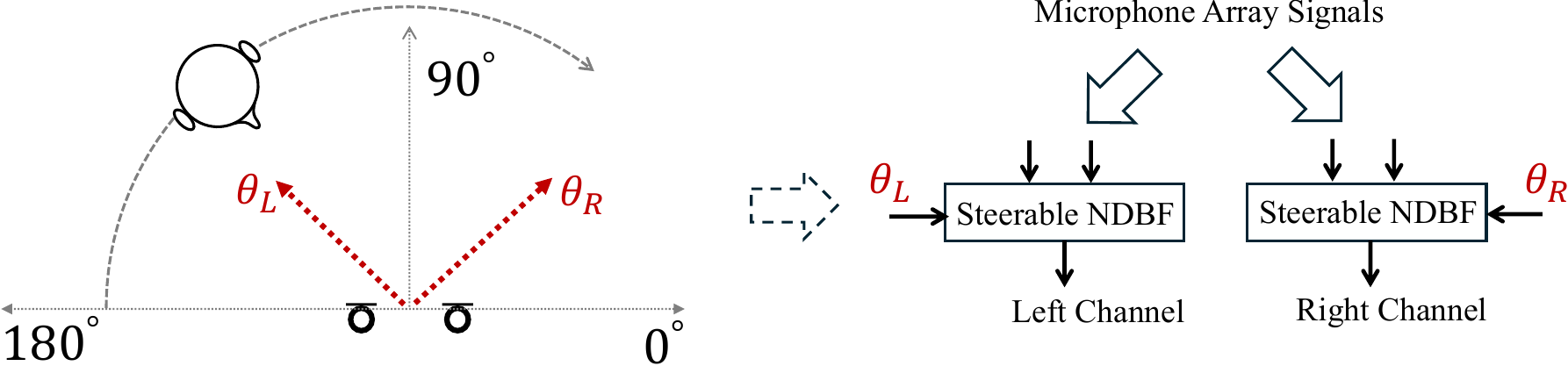}
	\caption{Stereo recording using two omnidirectional microphones by the proposed steerable NDBF.}
	\label{fig:stereo_recording}
\end{figure}

\begin{figure}[t!]
    \begin{minipage}[b]{0.429\linewidth} 
        \centering
        \includegraphics[width=\linewidth]{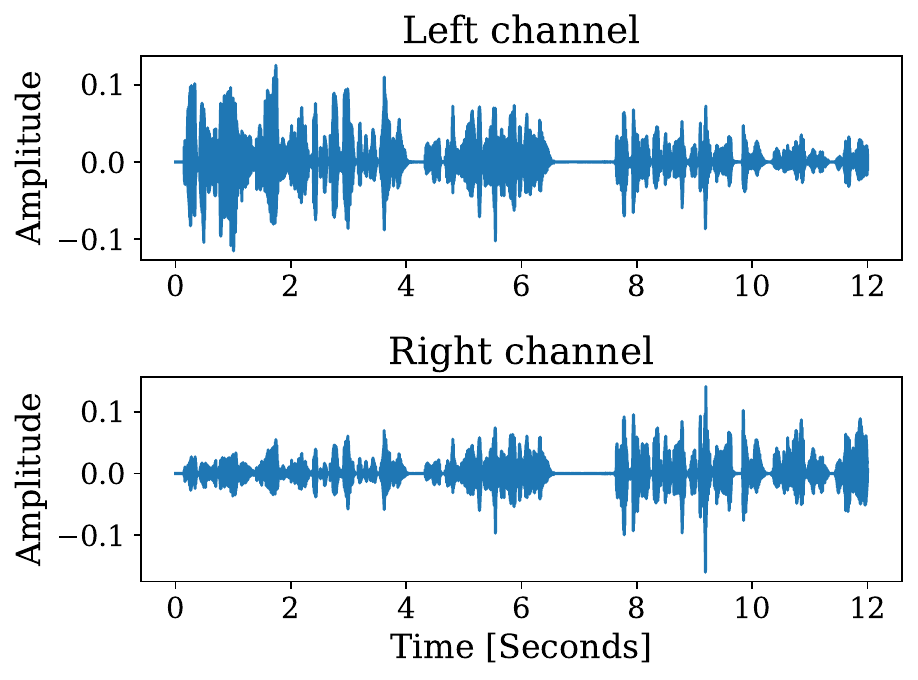}
        (a) NDBF stereo outputs.
    \end{minipage}
    \begin{minipage}[b]{0.429\linewidth} 
        \centering
        \includegraphics[width=\linewidth]{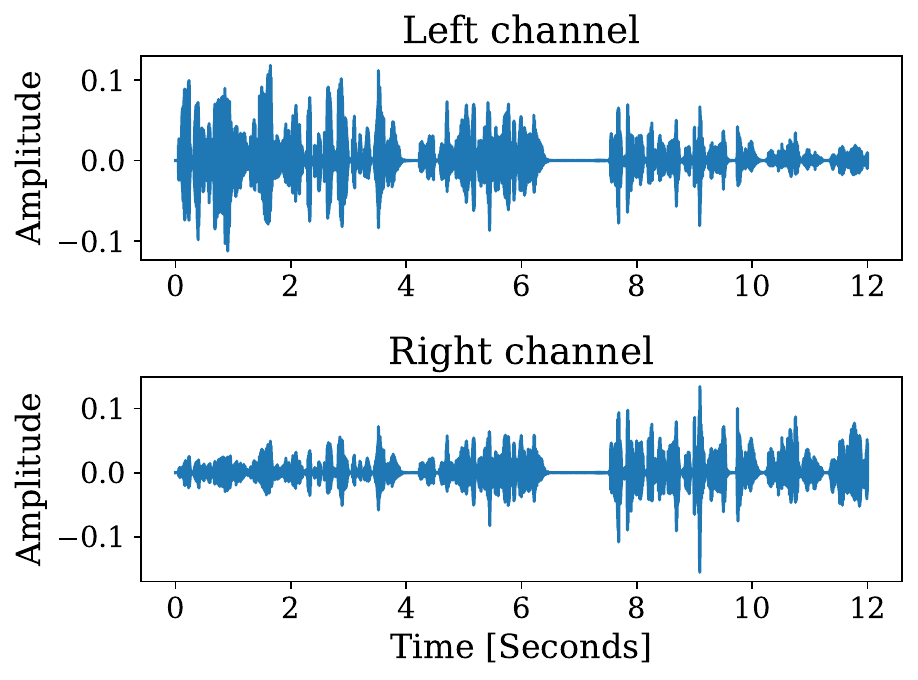}
        (b) VDM stereo outputs.
    \end{minipage}

        \begin{minipage}[b]{0.863\linewidth} 
        \centering
        \includegraphics[width=\linewidth]{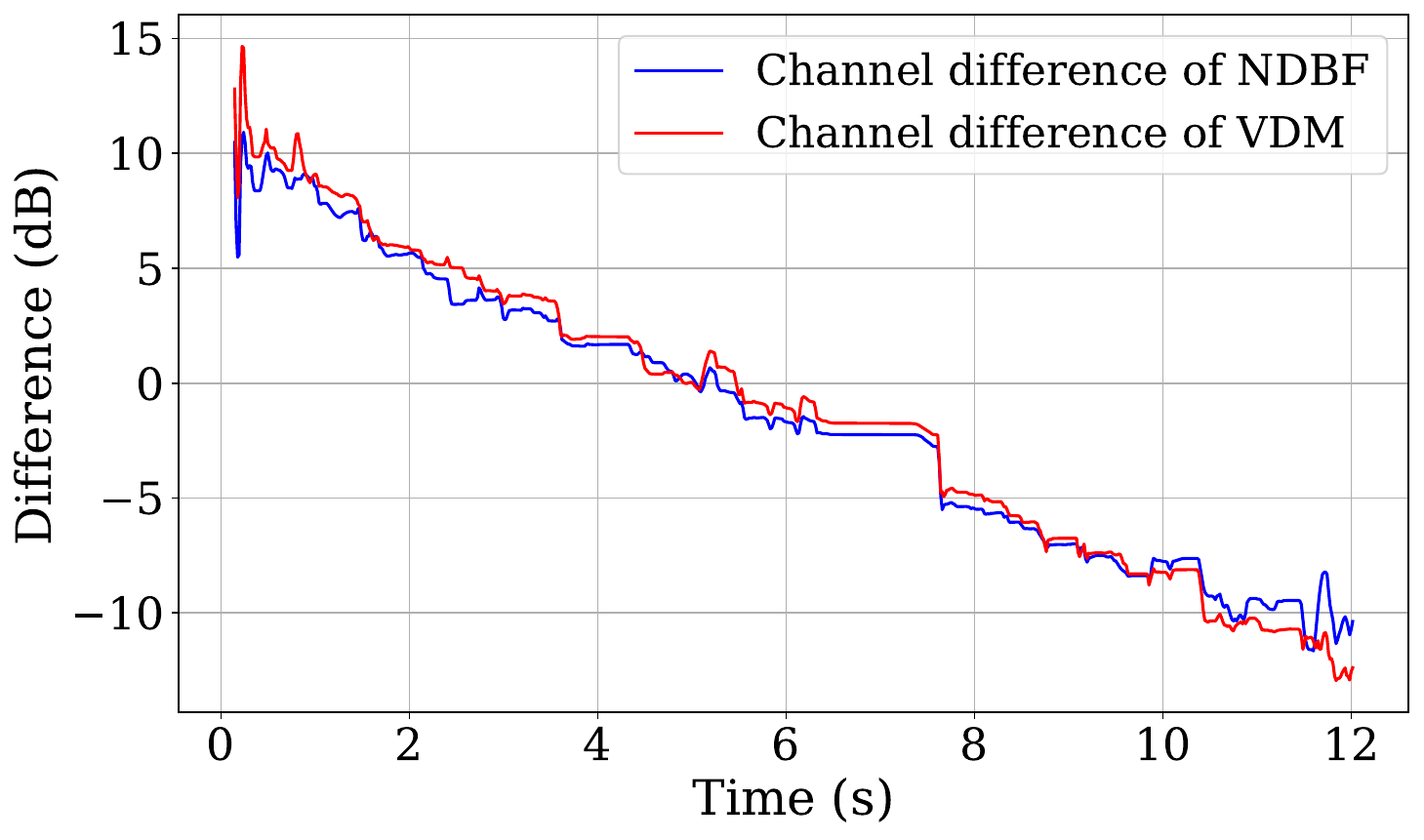}
        (c) Difference between the left and the right channel.
    \end{minipage}
    
	\caption{ Stereo output comparison between the proposed \ac{NDBF} and desired virtual directional microphones (VDM).} 
	\label{fig:stereo_diff}		
\end{figure}

We simulated a reverberant room with $\textrm{RT}_{60}$ of 0.15~\unit{\s} and a size of 6~\unit{\m} $\times$ 4~\unit{\m} $\times$ 3.5~\unit{\m}. As shown in the left side of Fig.~\ref{fig:stereo_recording}, a speech source moved steadily in a clockwise direction from $180^{\circ}$  to $0^{\circ}$ over approximately 12 seconds, with a fixed source-array distance of 1.5 m. The array consisted of two omnidirectional microphones spaced 3 cm apart. This simulated scene, generated using Dynamic Acoustic Scene Generator \cite{DASGenerator}, is used to evaluate \ac{NDBF}'s stereo recording capability.

To the best of our knowledge, neither classical differential beamformers nor existing neural beamformers enable stereo recording with only two closely spaced omnidirectional microphones. Traditionally, stereo recording using the X-Y technique requires two first-order directional microphones, most commonly each with a cardioid beampattern, arranged at look directions of $45^{\circ}$ and $135^{\circ}$ \cite{williams2002stereophonic}. In contrast, we trained a steerable \ac{NDBF} model with a first-order cardioid beampattern as the training target. The right side of Fig.~\ref{fig:stereo_recording} illustrates a configuration in which two-microphone signals are processed by two parallel steerable \ac{NDBF} models. One model is steered to $\theta_{\mathrm{L}} = 135^{\circ}$, while the other is steered to $\theta_{\mathrm{R}} = 45^{\circ}$. The resulting beamformed outputs are assigned to the left and right channels, respectively, thereby producing the stereo output of \ac{NDBF}, shown in Fig.~\ref{fig:stereo_diff}(a). For comparison, we simulate the X-Y technique using two \acp{VDM} located at the array center with the same look directions as the \ac{NDBF} models. The stereo output of \acp{VDM} is presented in Fig.~\ref{fig:stereo_diff}(b). This waveform comparison indicates that the stereo output of \ac{NDBF} closely matches that of \ac{VDM}. Figure~\ref{fig:stereo_diff}(c) presents the segmental energy differences between the left and right channels for \ac{NDBF} and \ac{VDM}; the two curves are quite close, indicating highly similar inter-channel level differences. This qualitative example demonstrates that \ac{NDBF}, even with a compact dual-microphone array, effectively captures the inter-channel level differences essential for stereo recording.

\section{Conclusions}
This paper presents a \ac{DNN}-based differential beamformer, termed NDBF, which achieves steerable high-order \ac{DMA} beampatterns using a dual-microphone array. Experimental results indicate that NDBF outperforms existing methods, particularly in approximating a target beampattern. This beampattern learning capability enables NDBF to overcome the limitations of traditional differential beamformers, which are typically first-order and non-steerable for such arrays. Furthermore, NDBF's steerability and beampattern-oriented spatial filtering enable stereo recording using only two closely spaced omnidirectional microphones. 

\pagebreak


\bibliographystyle{IEEEbib}
\bibliography{mybib}

\end{document}